\documentstyle[pre,aps]{revtex}
\tighten
\begin{document}


\title{\bf Area and Perimeter Distribution of a Surface in Two Dimensions \\ }
\author{\bf E.D. Moore\footnote{Electronic address: dmoore@thphys.ox.ac.uk}}
\address{Theoretical Physics, Department of Physics, Oxford
University,1 Keble Road, Oxford OX1 3NP, United Kingdom}
\date{\today}

\maketitle
\begin{abstract}
We consider the number of configurations of a surface in two dimensions that
has a prescribed length and encloses a prescribed perimeter with respect to a
baseline.  An approximate analytical treatment in a semi--continuum compares
favourably with results from an exact algorithm for the discrete lattice.  This
work is relevant for finding the entropy associated with macroscopic
configurations of such systems as domain growth problems,
evaporation--deposition problems, membrane physics, or polymer physics.
\end{abstract}

\pacs{PACS numbers: 05.50}
\narrowtext
\twocolumn

\section{Introduction}

In statistical mechanics, one very often performs averages over ensembles of
microscopic configurations.  Since this is usually very difficult, it is common
to replace the ensemble of microscopic quantities by an ensemble of macroscopic
quantities.  For instance, a spin Hamiltonian may be replaced by one that
depends only on magnetization.
This paper, we addresses the problem of a surface in two dimensions with fixed
endpoints where the macroscopic quantities are the perimeter of the surface and
the area enclosed by it.  Such a problem would arise, for example, if  one is
given a Hamiltonian for a system that consists of a area term and a surface
term, i.e.
\begin{equation}
{\cal H} = J f_s(S) - P f_a(A)
\end{equation}
where $S$ is the perimeter of the surface and $A$ is the area between it and
some substrate, $J$ is the surface tension, $P$ the pressure, and $f_s, f_a$
general functions of $S$ and $A$ respectively.
These types of Hamiltonians could arise in domain roughening
problems~\cite{cha81,bur81,kro81,chu81}, wetting problems,
evaporation--deposition problems, the physics of membranes, or polymer
physics~\cite{zwa68,lau70}.  We were motivated by an investigation of the
random field Ising model~\cite{intro,advanced}.

To find the free energy of such a system and thus any relevant macroscopic
quantities at a finite temperature, one would like to know the associated
entropy $S=k_B \ln \Omega_L(S,A)$.  Here, $\Omega_L(S,A)$ is the  number of
surface configurations with the given perimeter $S$ and enclosed area $S$ and
which begin and end at the ends of a line of length $L$; this is the quantity
that shall be of interest to us.

A slightly different version of this problem\footnote{They studied ``bar graph
polygons'' wherein the domain wall heights are not allowed to go below the
baseline.  In our model, the heights of the surface can be both positive and
negative.} has been solved exactly by Owczark and Prellberg~\cite{owc93} by
finding the generating function of $\Omega$.  However, the solution was in
terms of the $q$--series deformation of a hypergeometric function.  While
informative, this result is very complex and difficult to use in our analysis.
 Our aim is to find a simple form for $\Omega$ that can be easily manipulated.

We use an approximation to translate the problem into one of simple Riemannian
geometry that can be solved using the familiar tools of vector analysis.  The
result is the pleasingly elementary form
\begin{equation}
\label{eqn:simpleomega}
\Omega(S,A) = f \left( g^2 S^2 - \frac{A^2}{{\cal L}^2}\right)^\frac{L-3}{2}
\end{equation}
where $f$, ${\cal L}$ and $g$ are functions of $L$ only.  This result compares
favourably with exact numerical results.

\section{Definition of the Problem}
\label{sec:vaprob}
We consider a surface with fixed endpoints such that it has an end to end
distance of $L$.  The surface may take any path between these endpoints as long
as there are no overlaps or overhangs ({\it i.e.}\  it is a directed random
walk).  This should be a reasonable approximation in the limit of weak
disorder.  We label the straight line between the endpoints with the integer $i
\in [0,L]$ and the height of the surface at each point by $h_i$, a single
valued real number.  Figure~\ref{fig:va} illustrates these definitions.  The
quantities $A$ and $S$ may be written as functions of these
variables.
\begin{mathletters}
\label{eqn:VAI}
\begin{eqnarray}
\label{eqn:VA1}
	A&=& \sum_{i=0}^L h_i \\
\label{eqn:VA2}
     	S&=& \sum_{i=0}^L |h_i - h_{i-1}|
\end{eqnarray}
\end{mathletters}
where $S$ is the actual length of the surface less $L$ (since any surface
that covers the baseline distance must be at least $L$ long).

The problem is simplified somewhat by the introduction of the variables
$R_i = h_i - h_{i-1}$.  Like $h_i$, $R_i$ is a real number with $i \in [1,L]$
an
integer.  Under this change of variables, equation~(\ref{eqn:VAI}) becomes
\begin{mathletters}
\label{eqn:VAII}
\begin{eqnarray}
\label{eqn:VA3}
	A  &=& \sum_{i=1}^L (L-i) R_i  \\
\label{eqn:VA4}
	S  &=& \sum_{i=1}^L |R_i|.
\end{eqnarray}
\end{mathletters}

There is also an additional constraint that the surface finish at height
$h_L = 0$.  This introduces the auxiliary equation
\begin{equation}
\label{eqn:ht}
	\sum_{i=1}^L R_i = 0.
\end{equation}

Consider an ensemble of surfaces all with various $(A,S)$ and
subject only to the constraint~(\ref{eqn:ht}).
 In terms of the definitions above, we wish to know how many different values
of ${R_i}$ satisfy equations~(\ref{eqn:VA3}),~(\ref{eqn:VA4}),
and~(\ref{eqn:ht}) for fixed $A$ and $S$.
We solve this problem in the next section.

\section{Solution}

The solution is further complicated by the presence of the absolute value
function in~(\ref{eqn:VA4}) and so we will make a further approximation to
simplify matters.  It is hoped that this will not change the essential
character of the solution.
 Thus, we replace equation~(\ref{eqn:VA4}) by
\begin{equation}
\label{eqn:VA5}
g(L)^2 S^2 = \sum_{i=1}L R_i^2.
\end{equation}
This completes the approximations in the solution.  That  the form factor $g$
should only be a function of $L$ follows from a simple scaling argument.

Clearly, equation~(\ref{eqn:VA5}) implies that the vectors $R_i$ lie on a $L-1$
dimensional hypersphere embedded in $L$ dimensions.  The other two
equations~(\ref{eqn:VA3}) and~(\ref{eqn:ht}) introduce hyperplanes that pass
through the hypersphere.  Thus, vectors that satisfy all three equations lie on
the intersection
of the hypersphere with these two hyperplanes.  This intersection is itself a
hypersphere, but of
dimension $L-3$.  Thus, the problem is reduced to finding the measure of the
vectors which lie on the surface of this hypersphere.  The details of this
calculation are included in appendix~\ref{app:sol} and the result is
\begin{eqnarray}
\label{eqn:omfinal}
\Omega_L(S,A) &=&  f \left(\max\left( g^2 S^2 - \frac{A^2}{{\cal
L}^2},0\right)\right)^{\frac{L-3}{2}} \\
f &=& \pi^{\frac{L-2}{2}}{
\left(\frac{L-2}{2}\right)!}\left(L-2\right).\nonumber
\end{eqnarray}

It is worth noting that the problem posed by
equations~(\ref{eqn:VA5}),~(\ref{eqn:VA3}), and~(\ref{eqn:ht}) with $g(L)=1$ is
in itself worthy of consideration, and may relate to other models in
statistical physics.  However, since we were motivated by the random field
Ising model wherein~(\ref{eqn:VA4}) arises, we
 set $g(L)$ so that~(\ref{eqn:VA5}) maps as closely as possible back
to~(\ref{eqn:VA4}).  We describe how to do this in the following section.

\section{The Form Factor $\lowercase{g}(L)$}
\label{sec:g}

One can employ simple scaling arguments on~(\ref{eqn:omfinal}) to deduce that
$g(L) \sim c L^{-1/2}$ for large values of $L$, where we would expect $c$ to be
of order unity.  But, a better estimate of $g$ can be made.  Consider the set
of vectors that satisfy~(\ref{eqn:VA5}), that is,  the vectors on an $L$
dimensional hypersphere of radius $r= g(L) A$.  To satisfy~(\ref{eqn:VA4}) we
must set $A$ such that $\sum_{i=1}L R_i = A$ for all vectors on this
hypersphere.  Obviously, we cannot satisfy this criterion for all vectors, so
we demand that it be satisfied on average.  This presumes, of course, that all
the vectors on the hypersphere have equal importance in $\Omega_L$, which is a
reasonable first approximation.  Thus, we set
\begin{equation}
\langle \sum_{i=1}^{L}|x_i | \rangle = A
\end{equation}
where $\langle Q \rangle$ denotes the average value of a quantity $Q$ over the
hyperball of radius $r$.  It is a straightforward piece of mathematics to
perform this calculation, the details of which are included in
appendix~\ref{app:g}.  The result is
\begin{equation}
g(L) = \frac{\sqrt{\pi}}{L} \left( \frac{
\left(\frac{L+1}{2}\right)!}{\left(L/2\right)! \left(1 + 1/L\right) }
\right)
.\end{equation}
Using Stirling's approximation, we find that for large $L$ this tends
asymptotically to
\begin{equation}
g(L) \rightarrow \sqrt{\frac{\pi}{2 L}},
\end{equation}
which agrees with our earlier expectations.

\section{The Numerical Treatment}
\label{sec:num}

In order to test (and possibly improve upon) the analytic results obtained
above, an algorithm was devised to scan over all the random walks with
perimeter $S$ less than some preset maximum $A^\star$.  To understand this
algorithm, it is helpful to think of $S$ as the length of a piece of string
that we have to play with.  This piece of string must start and finish on the
baseline but may do anything in between (subject to the restriction that there
are no overhangs or overlaps).  But, the amount of string we have used up in
our previous steps determines how much we have left to play with in the
following steps.  For instance, at the first step, the string may have any
height $R_o \in [-A^\star/2, A^\star/2]$.    In general, if we are at position
$i$  and at height $h_i = \sum_{j=1}^{i-1} R_j$, and have already used up a
length of string $A_i = \sum_{j=1}^{i-1}|R_j|$, then this limits the range of
values $R_i$ may have.  Namely,
\begin{equation}
R_i \in \left[\frac{-A^\star + A_i - h}{2},\frac{A^\star - A_i - h}{2}\right]
.\end{equation}
Figure~\ref{fig:rvalues} illustrates this.

Furthermore, we only need to scan over various values of $R_i$ for $i \le L-1$
since $R_L$ is determined by the condition $\sum_{i=1}L R_i =0$.  A general
outline of the algorithm is given below:

{\sf
\begin{enumerate}
\item Set $R_1$ to $-A^\star/2$
\item Set all $R_i$ for $i=2,\ldots,L-1$ to their minimum possible value and
determine $R_L$
\item Measure $S$ and $A$.  Increase $\Omega(S,A)$ by one.
\item Increase $R_{L-1}$ by one.  If $R_{L-1}$ exceeds its maximum value then
increase $R_{L-2}$ by one and set $R_{L-1}$ to its minimum value.  If this
causes $R_{L-2}$ to exceed its maximum value, then increase $R_{L-3}$ by one
and set $R_{L-2}$ to its minimum value.  Et cetera.
\item Repeat steps 3 and 4 until $R_1 = A^\star/2$
\end{enumerate}
}

A recursive program was written in C to implement this algorithm, and the
program was run on Sun workstations.  Data was obtained for $A^\star$ as large
as 60 and for $L$ as large as 10.  It becomes increasingly difficult to get
data for larger $L$ since the number of random walks seems to grow
geometrically with $L$.  Indeed, we predict that
\begin{equation}
\int \Omega(S,A) {\rm d} V \sim (L-2)^L A^L \alpha^L
\end{equation}
where $\alpha$ is a constant of order unity.

\section{Comparison of Numerical and Analytic Results}
\label{sec:comp}

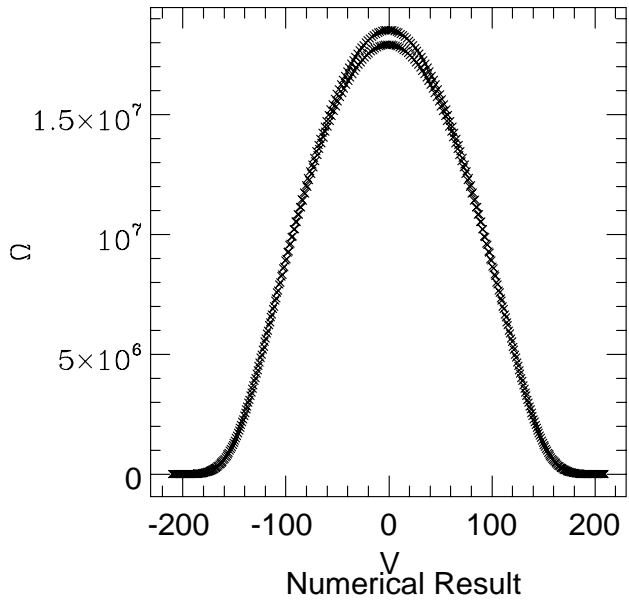
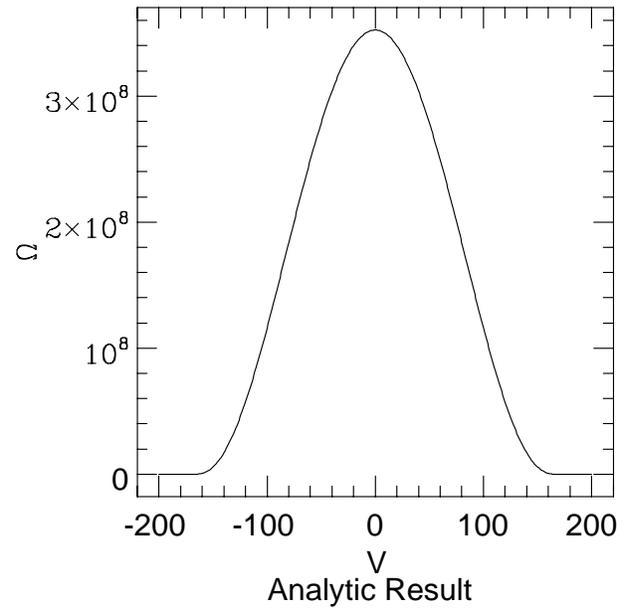
\begin{figure}
 \caption
{Sample data set from the numerical routine, compared
with the results of the analytic calculation}
 \label{fig:sampledata}
\end{figure}

When viewed side by side, the two results seem to look very
similar (see Figure~\ref{fig:sampledata}).  However, the scales do not
agree.  This is attributed to the fact that the numerical routine
 investigates only discrete values of $R_i$ whereas in the
analytic treatment $R_i$ was allowed to be continuous.
In both treatments $L$ is divided into discrete intervals, so this is
not a cause for worry (as long as we have $L > 3$).
{}From the point of view of our hypersphere argument, the analytic
treatment has given the measure of all vectors on the surface,
whereas if the components of the vectors are constrained to be
integers, there are fewer vectors close to the surface of the
hypersphere.  Nowhere is the effect of the discretization more
noticeable than in the fact that $\Omega(S,A)=0$ on a discrete lattice for odd
$S$.
This simply says that we need as many steps up as down if we are
to return to the origin.  We do not have this same constraint in
the semi--continuous model.
Nevertheless, the results still compare very favourably on various
 levels (up to this scaling factor).  Three quantities were chosen
 for the purposes of comparison, which we list below together with their
predicted values:
\begin{itemize}
\item The maximum value of $\Omega$ for a given $S$ and $L$,
\begin{equation}
M_m=\Omega_L(A,0) = f(L) \left( g(L) A\right)^{L-3}.
\end{equation}
\item The zeroth moment of the distribution with respect to $A$,
\begin{equation}
M_o = \int \Omega(S,A){\rm d} V = {\cal L} f(L) \left( g(L) A \right)^{L-2}
\Upsilon(L)
\end{equation}
 where
\begin{equation}
\Upsilon(L) = \frac{2^{2L-5} (L-3)! (L-3)!}{(2L-5)(2L-6)!}.
\end{equation}
\item The second moment of the distribution with respect to $A$,
\begin{equation}
M_2 = \int \Omega(S,A) A^2 {\rm d} V = {\cal L}^3 f(L) \left(g(L) A\right)^L
\frac{\Upsilon(L)}{2L-3}.
\end{equation}
\end{itemize}

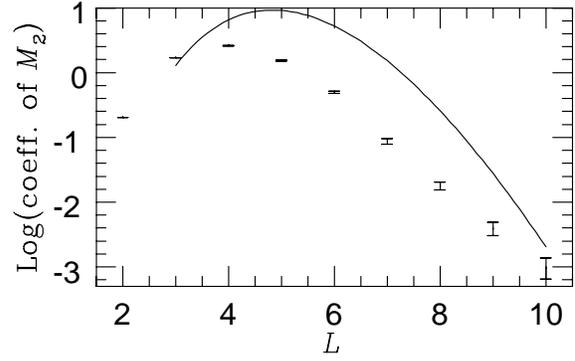
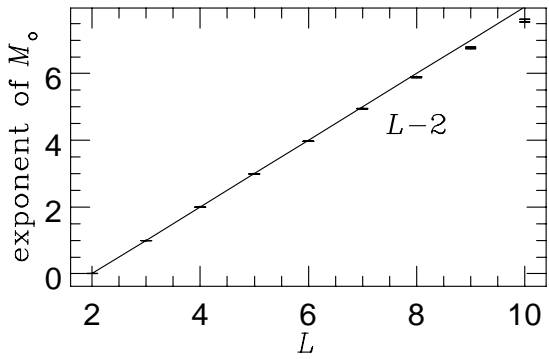
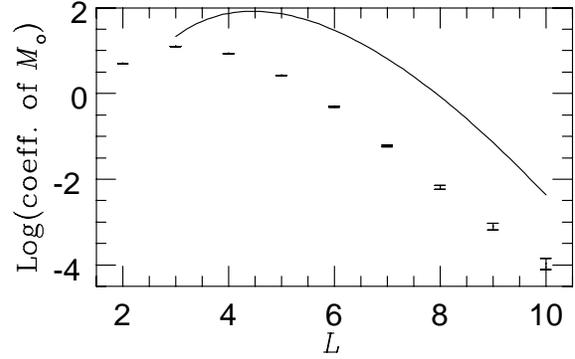
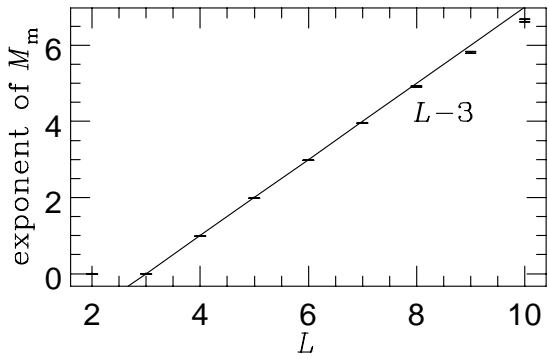
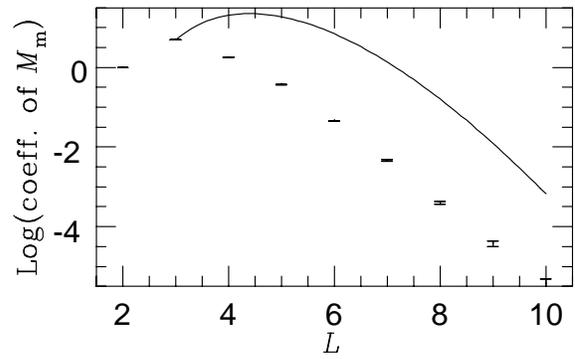
\begin{figure}
\caption
{Plots of the exponent  and coefficient of $M_m$, $M_0$ and $M_2$.  The data
points are those values found from numerical work while the solid lines
represent the theoretical prediction.  The error bars represent the statistical
error from the power law fit.}
\label{fig:momentplot}
\end{figure}

Given that we expect these forms to arise, we generate these
quantities from our data and fit them to a power law behaviour
 in $S$ for various values of $L$.  The exponent and coefficient of this fit
are then compared to the analytic predictions listed above.  The results are
shown in Figure~\ref{fig:momentplot}, where one can see that the exponents, but
not the coefficients, agree well (as expected from the above discussion).
However, the coefficients do seem to have the correct scaling behaviour in $L$.

\section{Conclusions}
\label{sec:conc}

We have investigated the question of the entropy of a two dimensional surface
with fixed volume and area from both the numerical and analytic viewpoints.
{}From the analytic side, we found an elegant expression for the
semi--continuum by using a single approximation.  From the numerical side, we
generated exact results that compared favourably with the analytic expression.
However, there were certain differences noted between the two and these were
attributed to differences between the discrete and semi--continuous
formulations of the problem.

Nevertheless, the solution captures the important characteristics of the
numerical work.  It is hoped that this work will be of value to finite
temperature formulations of problems in many areas of statistical mechanics
where the question may be posed in terms of such macroscopic quantities as the
length of a curve and the area enclosed by it.

\acknowledgements

The author would like to thank the Rhodes Trust and the Natural Sciences and
Research Council of Canada for providing funding for this work.  Also, I would
like to thank Dr. R.B. Stinchcombe for many useful discussions on this topic.

\appendix

\section{Calculation of $\Omega(A,V)$}
\label{app:sol}

We wish to find the measure of all vectors that satisfy
equations~(\ref{eqn:VA3}),~(\ref{eqn:ht}), and~(\ref{eqn:VA5}).
To simplify matters, let us introduce the dimensionless variables $r_i^\prime =
R_i/(g A)$. Working with these vectors instead of $R_i$ introduces a change of
measure and since the object we are looking at is clearly a $L-3$ dimensional
hypersphere, this change of measure is $J= (g A)^{L-3}$.  Written in these new
coordinates the equations~(\ref{eqn:VA3}),~(\ref{eqn:ht}), and~(\ref{eqn:VA5})
become
\begin{mathletters}
\label{eqn:master}
\begin{eqnarray}
\sum_{i=1}^L r_i^{\prime 2} &=& 1,  \\
\sum_{i=1}^L r_i ^\prime &=& 0, \\
\sum_{i=1}^L (L-i+1) r_i^\prime &=& v,
\end{eqnarray}
\end{mathletters}
where $v = V/g(L) A$.

This change of measure associated with the change of variables is associated
with the scaling behaviour of the equations.  We can make this more explicit by
writing
\begin{equation}
\label{eqn:scaling1}
\Omega_L(A,V) = J \Omega_L(1/g(L),V/g(L) A)
.\end{equation}
Similarly, we could have written
\begin{equation}
\label{eqn:scaling2}
\Omega_L(A,V) = V^{L-3} \Omega_L(A/V,1)
.\end{equation}
It is important that our final solution obey these two relations.

The path to the solution is made clearer if we rewrite the equations in
{}~(\ref{eqn:master}) so that they more closely resemble the vector fomalism.
To achieve this, let us introduce the vectors
\begin{mathletters}
\begin{eqnarray}
\vec{r}^\prime &=& \{r_1^\prime,r_2^\prime,\ldots,r_L^\prime\} \\
\vec{\openone}^\prime &=& \{1,1,\ldots,1\} \\
\vec{L}^\prime &=& \{L-1,L-2,\ldots,0\}
\end{eqnarray}
\end{mathletters}
so that~(\ref{eqn:master}) becomes
\begin{mathletters}
\begin{eqnarray}
\label{eqn:vecsphere}
\|\vec{r}^\prime\| &=& 1 \\
\label{eqn:ortho}
 \vec{r}^\prime \cdot \vec{\openone}^\prime &=& 0 \\
\label{eqn:dot}
 \vec{r}^\prime\cdot\vec{L}^\prime &=& \|\vec{L}^\prime\| \cos \theta = v
\end{eqnarray}
\end{mathletters}
where $\|\vec{L}^\prime\|^2 = L (L-1) (2L-1)/6$.

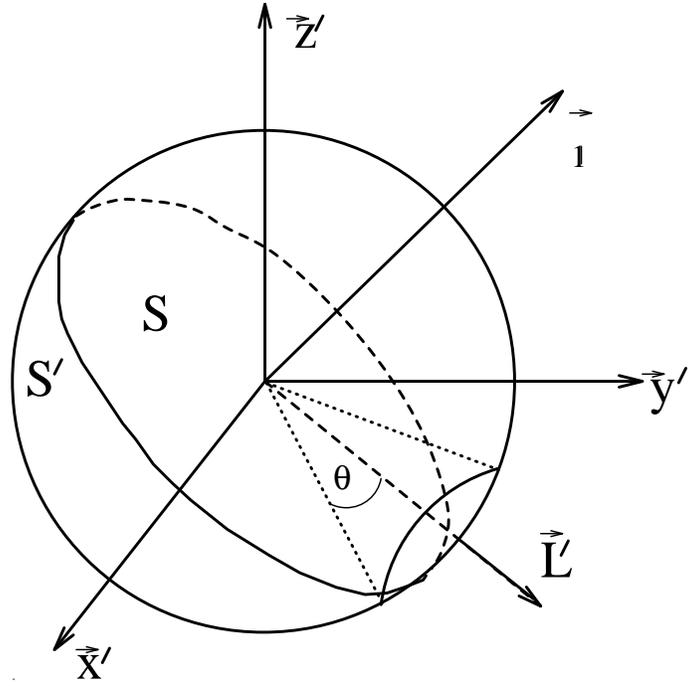
\begin{figure}
   \caption[]{The vector space that $\vec{\openone}$ and $\vec{L}^\prime$ live
in, showing the hyperspheres $S$ and $S^\prime$.}
   \label{fig:hsphere}
\end{figure}

Figure~\ref{fig:hsphere} gives a pictorial representation of these equations.
Equation~(\ref{eqn:vecsphere}) specifies that the vectors $\vec{r}^\prime$ lie
on a unit hypersphere, $S^\prime$, of dimension $L-1$.
Equation~(\ref{eqn:ortho}) is an orthogonality condition and is used to lower
the dimension of the problem by one.  Finally, equation~(\ref{eqn:dot})
specifies that the vectors are at an angle $\theta$ to the vector
$\vec{L}^\prime$.

We will now employ a coordinate rotation and project our problem onto the $L-1$
dimensional space othogonal to $\vec{\openone}$.  The rotation is chosen to
simplify the projection, {\it e.g.}\  we let the $\hat{z}$ axis lie parallel to
the vector $\vec{\openone}$ so that $\vec{r}^\prime \rightarrow (\vec{r},0)$
and $\vec{L}^\prime \rightarrow (\vec{L}, L_z)$.

Now $\vec{r}$ and $\vec{L}$ are both $L-1$ dimensional vectors and  $L_z$ is
the component of $\vec{L}^\prime$ parallel to $\vec{\openone}$.
\begin{equation}
L_z = \frac{\vec{L}^\prime \cdot \vec{\openone}}{\|\vec{\openone}\|} =
\frac{L(L-1)}{2 \sqrt{L}}
.\end{equation}
Equation~(\ref{eqn:dot}) retains its familiar form in this new, unprimed,
coordinate system
\begin{equation}\label{eqn:phi}
\vec{r}\cdot\vec{L} = \|\vec{L}\| \cos \theta = v
.\end{equation}

We want a measure of the set of all unit vectors $\vec{r}$ that satisfy this
dot product.  These vectors form a hypercone with angle of opening $2\theta$
and sides of unit length.  An appropriate measure is the area of the $L-2$
dimensional hypersphere that the tips of these vectors trace out.  In general,
the area of a $d$ dimensional hypersphere of radius $r$ is
\begin{equation}
A_d(r) = \frac{\pi^{d/2}d}{(d/2)!} r^{d-1}
.\end{equation}
Letting $d \rightarrow L-2$ and $ r \rightarrow \sin \theta$ we arrive at
\begin{equation}
\label{eqn:omalmost}
\Omega(1,v) = \pi^{\frac{L-2}{2}}{ \left(\frac{L-2}{2}\right)!}\left(L-2\right)
(1-\cos^2\theta)^{\frac{L-3}{2}}
.\end{equation}
We substitute for the value of $\cos \theta$ using equation~(\ref{eqn:phi}) and
simplify the notation by letting ${\cal L} = \| \vec{L}\|$, i.e.
\begin{eqnarray}
{\cal L}^2 &=& \|\vec{L}^\prime\|^2 - L_z^2 \nonumber\\
           &=& \frac{L(L-1)(2L-1)}{6} - \frac{L(L-1)^2}{4} \\
\label{eqn:cos}
\cos \theta &=& \frac{v}{{\cal L}}.
\end{eqnarray}
Using the scaling relation~(\ref{eqn:scaling1}) we arrive at our final form
\begin{eqnarray}
\label{eqn:omfinal2}
\Omega_L(A,V) &=& f \left( g^2(L) A^2 - \frac{V^2}{{\cal
L}^2}\right)^{\frac{L-3}{2}} \\
f &=&  \pi^{\frac{L-2}{2}}{ \left(\frac{L-2}{2}\right)!}\left(L-2\right)
\nonumber
.\end{eqnarray}

It is simple to check that this form satisfies the scaling
relations~(\ref{eqn:scaling1}) and~(\ref{eqn:scaling2}).  Obviously, for
physical reasons, we set $\Omega(A,V) = 0$ if
$V > g(L) A {\cal L}$.

\section{Calculation of $\lowercase{g}(L)$}
\label{app:g}

Following the philosophy outlined in section~\ref{sec:g} we wish to find the
sum of the absolute values of the components of a vector that lies on an $L$
dimensional hyphersphere of radius $r$.  Written in terms of spherical polar
coordinates, these components are
\begin{eqnarray}
\label{eqn:x}
x_1 &=& r \cos\theta_1 \nonumber\\
x_2 &=& r \sin\theta_1\cos\theta_2 \nonumber\\
\vdots \nonumber \\
x_i &=& r \prod_{j=1}^{i-1} \sin\theta_j\cos\theta_i \nonumber\\
\vdots \nonumber \\
x_L &=& r \prod_{j=1}^{L-1} \sin\theta_j
.\end{eqnarray}

We wish to consider $\langle \sum_{i=1}^L |x_i| \rangle$ but we can take
advantage of the spherical symmetry to write this as $L \langle x_L\rangle$.
All that remains is to tackle the integral
\begin{equation}I = \oint \sqrt{G} {\rm d}\theta_1{\rm d}\theta_2\ldots{\rm
d}\theta_{L-1}\left|\prod_{i=1}^{L-1}\sin\theta_i\right|
\end{equation}
where $G=r^{2(L-1)} \prod_{i=1}^{L-2} \sin^{2(L-i-1)}\theta_i$ is the metric of
the surface of this hypersphere.  Expanding out the integrals,
\begin{equation}I = 2 \prod_{i=1}^{L-2}\int_0^\pi \sin\theta_i
\sin^{L-i-1}\theta_i{\rm d}\theta_i  \int_0^{\pi}\sin\theta_{L-1}{\rm
d}\theta_{L-1}
.\end{equation}
The remaining integral may be done by comparison to the area integral of the
$d$ dimensional unit sphere:
\begin{equation}A_o(d) = 2\pi \prod_{i=1}^{d-2} \int_0^\pi
\sin^{d-i-1}\theta_i{\rm d}\theta_i = \frac{\pi^{d/2}d}{(d/2)!}
\end{equation}
since
\begin{equation}I = 2\prod_{i=1}^{L-1} \int_0^{L-i}\theta_i{\rm d}\theta_i =
\frac{A_o(L+1)}{\pi}
.\end{equation}
Thus, the normalized average is
$
\langle \sum_{i=1}^L |x_i| \rangle = L I/A_o(L)
 \equiv A
$.  Putting $r=A g(L)$ we get the final form
\begin{equation}g(L) = \frac{\sqrt{\pi}}{L} \left( \frac{
\left(\frac{L+1}{2}\right)!}{\left(L/2\right)! \left(1 + 1/L\right) }
\right)
.\end{equation}


\begin{figure}
   \caption{Directed surface with area $S$ and perimeter $S$.}
   \label{fig:va}
\end{figure}

\begin{figure}
\caption
{ The drawing on the left shows the maximum value $R_o$
can have, thus using up all the string.  The drawing on the left
show the maximum value $R_i$ can have, so that we use up all the
available string at this step.}
\label{fig:rvalues}
\end{figure}

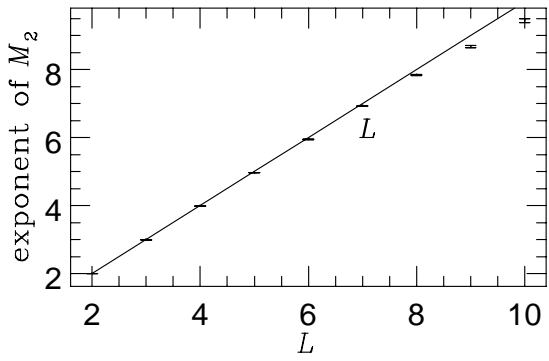
\end{document}